\documentclass[runningheads,a4paper]{llncs}

\usepackage{amssymb}
\usepackage{graphicx}

\pdfoutput=1

\usepackage{url}
\urldef{\mailsa}\path {robledo@fisica.unam.mx}
\newcommand{\keywords}[1]{\par\addvspace\baselineskip
\noindent\keywordname\enspace\ignorespaces#1}

\begin{document}

\mainmatter  

\title{Possible thermodynamic structure underlying the laws of Zipf and Benford}

\titlerunning{Possible thermodynamic structure underlying the laws of Zipf and Benford}


\author{Carlo Altamirano\inst{1} \and Alberto Robledo\inst{1}, \inst{2}}

\authorrunning{Possible thermodynamic structure underlying the laws of Zipf and Benford}

\institute{Instituto de F\'{i}sica, Universidad Nacional Aut\'onoma de M\'exico,\\
Apartado postal $20-364$, M\'{e}xico $01000$ D.F., Mexico\\
\and Grupo Interdisciplinar de Sistemas Complejos, Departamento de Matem\'{a}ticas,
\\ Universidad Carlos III de Madrid, $28911$ Legan\'{e}s, Madrid, Spain}


\toctitle{Possible thermodynamic structure underlying the laws of Zipf and Benford}

\maketitle

\begin{abstract}
We show that the laws of Zipf and Benford, obeyed by scores of numerical data generated by many and diverse kinds of natural phenomena and human activity are related to the focal expression of a generalized thermodynamic structure. This structure is obtained from a deformed type of statistical mechanics that arises when configurational phase space is incompletely visited in a severe way. Specifically, the restriction is that the accessible fraction of this space has fractal properties. The focal expression is an (incomplete) Legendre transform between two entropy (or Massieu) potentials that when particularized to first digits leads to a previously existing generalization of Benford's law. The inverse functional of this expression leads to Zipf's law; but it naturally includes the bends or tails observed in real data for small and large rank. Remarkably, we find that the entire problem is analogous to the transition to chaos via intermittency exhibited by low-dimensional nonlinear maps. Our results also explain the generic form of the degree distribution of scale-free networks.
\end{abstract}

\section{Introduction}
\label{intro}

Over more than half a century, observers of the astonishing ubiquity of the
empirical laws of Zipf and Benford have been puzzled by their seeming
universal validity and intrigued about the plausible answer to the central
question of why they appear in so many contexts. As it is widely known,
Zipf's law refers to the (approximate) power law that is displayed by sets
of data (populations of cities, words in texts, impact factors of scientific
journals, etc.) when these are given a ranking (in relation to size of
populations, frequency of words, magnitude of impact factors, etc.) \cite%
{zipf1}. Benford's law is a well-known simple logarithmic rule for the
frequency of first digits found in listings of data (stock market prices,
census data, heat capacities of chemicals, etc.) \cite{benford1}.

It has been argued that Benford's law is a special case of Zipf's law \cite%
{precursor1}. Indeed the relationship between the two has been made explicit
some years ago \cite{pietronero1} by first obtaining a generalization of
Benford's law from the basic assumption that the underlying probability
distribution $P(N)$ of the data $N$ under consideration is scale invariant
and therefore has the form of the power law $P(N)\sim N^{-\alpha }$, $\alpha
\geq 1$. A simple integration over $P(N)$, to obtain the relative
probability for consecutive integers $n$ and $n+1$, leads, when $\alpha =1$,
to $\pi (n)=\log (1+n^{-1})$ which is Benford's law. The next step in Ref. 
\cite{pietronero1} was to obtain the rank $k$ from $P(N)$, this time as an
integration over $P(N)$ from $N(k)$, the number of data that define the rank 
$k$, to a finite number $N_{\max }$ that corresponds to the first value of
the rank $k$. In the limit $N_{\max }\rightarrow \infty $ they obtain $%
N(k)\sim k^{1/(1-\alpha )}$ that is Zipf's law with exponent $1/(1-\alpha )$
when $\alpha >1$.\ For many sets of real data $\alpha \simeq 2$ and the
standard Zipf law is $\alpha =2$ \cite{pietronero1}.

Here we expand on the results of Ref. \cite{pietronero1}. Our first, simple,
step is to keep $N_{\max }$ finite, but as we argue below, this
consideration facilitates the articulation of a major inference on the
physical nature of the laws of Zipf and Benford. We contend that these laws
are related to a general thermodynamic expression, albeit for a special type
of thermodynamic structure obtained from the usual via a scalar deformation
parameter represented by the power $\alpha $. The general thermodynamic
expression is seen to represent an (incomplete) Legendre transform (akin to
a Landau free energy or a free energy density functional) between two
thermodynamic potentials, the expression relating the corresponding
partition functions becomes a generalized Zipf's law. We identify these
quantities as well as the conjugate variables involved, which are the rank $%
k $ and the inverse of the total number of data $\mathcal{N}^{-1}$. We also
reason that this kind of deformed thermodynamics arises from the existence
of a strong impediment in accessing configurational phase space, that
materializes in only a fractal or multifractal subset of this space being
available to the system. A quantitative consequence of considering $N_{\max
} $ finite is the reproduction of the small-rank bend displayed by real data
before the power-law behavior sets in. The power law regime in the
theoretical expression persists up to infinite rank $k\rightarrow \infty $,
indicating a kind of `thermodynamic limit'. We illustrate this feature by
comparison with on hand data for frequencies of English words in texts \cite%
{words1}. We refer to the application of this scheme to the degree
distribution of scale-free networks.

A subsequent development is the identification of a strict analogy between
the aforementioned thermodynamic expression and that for (all) the
trajectories at the transition to chaos via intermittency in nonlinear
low-dimensional maps, the so-called tangent bifurcation \cite{schuster1}.
These critical trajectories follow \cite{baldovin1} the exact closed form of
the functional-composition Renormalization Group (RG) fixed-point map \cite%
{hu1} \cite{schuster1}. Consequently, we associate the same
statistical-mechanical structure to the nonlinear dynamics at this
transition. Further, we examine the modifications brought upon the
generalized law of Zipf by the corresponding shift of the map out of
tangency into the chaotic regime. These consist in the introduction of an
upper bound for the rank $k$ and the reproduction of the tail observed for
large rank in real data. We illustrate our scheme by comparison with
available numerical data for the so-called eigenfactor of physics journals 
\cite{eigenfactor1}, industrial production rates \cite{industrial1}, and
carbon emissions \cite{emissions1}. The analogy indicates that the most
common value for the index $\alpha $ should be $\alpha =2$.

Lastly, we make use of the statistical-mechanical interpretation to extend
our analysis. We presuppose that the Legendre transformation expressed by
these laws can be finalized in the usual way and eliminate the variable $%
\mathcal{N}^{-1}$ in favor of $k$. To accomplish this step it is necessary
to specify the (partition) function $N_{\max }(\mathcal{N}^{-1})$, a feature
of the available data or a prerogative of the data collector, and evaluate
the `equation of state' $k(\mathcal{N}^{-1})$. In doing this it becomes
clear that the universality of the laws is due to the general form of the
incomplete Legendre transformation, while the initial and transformed
thermodynamic potentials are particular of the data in hand.

Thus, in the next Section 2 we reproduce the expressions in Ref. \cite%
{pietronero1} relevant to our purposes. In Section 3 we describe the
generalized statistical-mechanical structure we observe in these
expressions. In the following Section 4 we present the parallelism between
the ranking of data and the critical dynamics at the tangent bifurcation in
nonlinear maps and describe the finite size effect of the former. In Section
5 we extend the statistical-mechanical description and draw conclusions on
the apparent universality of the aforementioned empirical laws. We conclude
in Section 6 with a summary and discussion. A partial preliminary account of
the contents of this paper appeared in Ref. \cite{altamirano1}

\section{Derivation of the Laws of Benford and Zipf}
\label{derivation}

Denote by $P(N)$ the probability distribution associated to the set of data
under consideration (e.g., the distribution obtained from a histogram
generated by data - a total of $\mathcal{N}$ numbers - giving the magnitudes
of the population of a set of countries). Under the assumption of scale
invariance the distribution has the form of a power law $P(N)\sim N^{-\alpha
}$, $\alpha >0$. The probability of observation of the first digit $n$ of
the number $N$ is given by \cite{pietronero1}%
\begin{equation}
\pi (n)=\int\limits_{n}^{n+1}N^{-\alpha }dN=\frac{1}{1-\alpha }\left[
(n+1)^{1-\alpha }-n^{1-\alpha }\right] ,  \label{benford1}
\end{equation}%
$\alpha \neq 1$, from which one obtains Benford's law $\pi (n)=\log
(1+n^{-1})$ when $\alpha =1$.

The set of $\mathcal{N}$ factual data numbers can be ranked and compared
with ranking of another set of also $\mathcal{N}$ numbers extracted from the
basic distribution $P(N)\sim N^{-\alpha }$. The rank $k$ is given by $k=%
\mathcal{N\ }\Pi \mathcal{(}N(k),N_{\max })$ where \cite{pietronero1}%
\begin{eqnarray}
\Pi (N(k),N_{\max }) &=&\int\limits_{N(k)}^{N_{\max }}N^{-\alpha }dN 
\nonumber \\
&=&\frac{1}{1-\alpha }\left[ N_{\max }^{1-\alpha }-N(k)^{1-\alpha }\right] ,
\label{rank1}
\end{eqnarray}%
$\alpha \neq 1$, where $N_{\max }$ and $N(k)$ correspond, respectively, to
rank $k=0$, and nonspecific rank $k>0$. Solving the above for $N(k)$ in the
limit $N_{\max }\gg 1$ yields Zipf's law $N(k)\sim k^{1/(1-\alpha )}$. Eq. (%
\ref{rank1}) introduces a continuum-space variable for the rank $k$ in which
the first value of the rank is $k=0$. This is a departure from the usual
representation with first rank $k=1$ and the following ranks given by
successive natural numbers. This approach corresponds to a continuum
variable description suitable for large data sets, and for which restriction
to integer values of the rank can be obtained by use of suitable values for
the lower limits of integration $N(k)$ in Eq. (\ref{rank1}).

\section{Generalized laws of Benford and Zipf as thermodynamic relations}
\label{generalized}

Consider the $q$-deformed logarithmic function $\log _{q}(x)\equiv
(1-q)^{-1}[x^{1-q}-1] $ with $q\neq 1$ a real number, and its inverse, the $%
q $-deformed exponential function $\exp _{q}(x)\equiv \left[ 1+(1-q)x\right]
^{1/(1-q)}$ that reduce, respectively, to the ordinary logarithmic and
exponential functions when $q=1$. In terms of these functions, Eq. (\ref%
{rank1}) and its inverse can be written more economically as

\begin{equation}
\log _{\alpha }N(k)=\log _{\alpha }N_{\max }-\mathcal{N}^{-1}k,
\label{benford2}
\end{equation}%
and%
\begin{equation}
N(k)=N_{\max }\exp _{\alpha }[-N_{\max }^{\alpha -1}\mathcal{N}^{-1}k].
\label{zipf2}
\end{equation}%
We first comment that Eq. (\ref{zipf2}) is a generalization of Zipf's law
that takes properly into account the behavior for low rank $k$ observed in
real data where, as one would expect, $N_{\max }$ is finite. In Fig.~\ref%
{fig:words} we compare the numbers of occurrences of English words in a
corpus with $N(k)$ as given by Eq. (\ref{zipf2}) where the reproduction of
the small-rank bend displayed by the data before the power-law behavior sets
in is evident. In the theoretical expression this regime persists up to
infinite rank $k\rightarrow \infty $. Alternatively, we recover from Eq. (%
\ref{zipf2}) the power law $N(k)\sim k^{1/(1-\alpha )}$ in the limit $%
N_{\max }\gg 1$ when $\alpha >1$. We note that for ranked listings of data $%
N $ the normalization of their distribution $P(N)$ implies that the maximum
rank $k_{\max }$ is equal to the number of data $\mathcal{N}$. Normalization
of $P(N)=$ $N^{-\alpha }$ leads to $k_{\max }=$ $\mathcal{N}$ \ with both $%
k_{\max }\rightarrow \infty $ and $\mathcal{N}\rightarrow \infty $, but $%
\mathcal{N}^{-1}k$ generally finite. The assumption of a pure power law form
for $P(N)$ cannot represent a set with a finite number of data.

\begin{figure}[tbp]
\centering
\includegraphics[height=6.2cm]{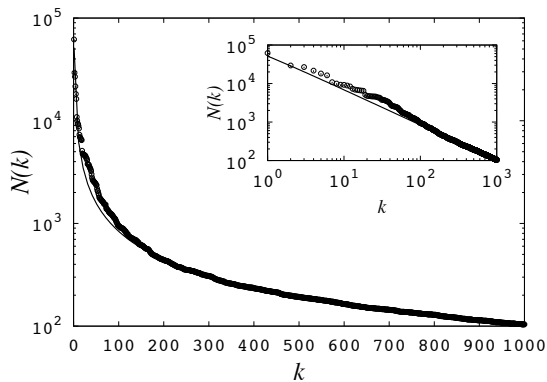}
\caption{Rank-order statistics for the occurrence of words (\emph{empty
circles}) in the British National Corpus \protect\cite{words1}. Eq. (\protect
\ref{zipf2}) with $\protect\alpha =2.09$ (\emph{smooth curve}) is fitted to
the data. The straight line in the inset is drawn for visualization
purposes. }
\label{fig:words}
\end{figure}

Now, in order to arrive at an interesting physical interpretation of Eq. (%
\ref{benford2}) we look at the quantities contained in it. We notice that
both $\log _{\alpha }N_{\max }$ and $\log _{\alpha }N(k)$ are given by the
integrals%
\begin{equation}
\log _{\alpha }N_{\max }=\int\limits_{1}^{N_{\max }}N^{-\alpha }dN\;
\textnormal{and} \;\log _{\alpha
}N(k)=\int\limits_{1}^{N(k)}N^{-\alpha }dN,  \label{integrals1}
\end{equation}%
and these in turn can be seen, when $\alpha =1$, to conform to the
evaluation of entropy $\widehat{S}_{1}=\log N_{\max }$ or $S_{1}=\log N(k)$
where the probability of $N$ equally-probable configurations in phase space
is $P(N)=$ $N^{-1}$. If we now allow for $\alpha >1$ we can retain the same
interpretation,%
\begin{equation}
\widehat{S}_{\alpha }=\log _{\alpha }N_{\max }\; \textnormal{and}%
\;S_{\alpha }=\log _{\alpha }N(k),  \label{entropies1}
\end{equation}%
with $P(N)=$ $N^{-\alpha }$ still viewed as the probability of $N$
equally-probable phase-space configurations, and with $N_{\max }$ and $N(k)$
playing the roles of total configurational numbers or partition functions.
Therefore Eq. (\ref{benford2}) can be rewritten as%
\begin{equation}
S_{\alpha }=\widehat{S}_{\alpha }-\mathcal{N}^{-1}k,  \label{legendre1}
\end{equation}%
and read as the expression of what we refer to as an \emph{incomplete}
Legendre transform from the Massieu potential $\widehat{S}_{\alpha }(%
\mathcal{N}^{-1})$, a function of the inverse of the number $\mathcal{N}$,
to the entropy $S_{\alpha }(k)$, a function of the rank $k$. The conjugate
variables $\mathcal{N}^{-1}$ and $k$ could be seen to play the roles, for
example, of inverse temperature $\beta $ and energy $u$ in the description
of a thermal system. As we know the Legendre transform is performed in two
steps, the first is to add (subtract) the product of two conjugate variables
from one thermodynamic potential and the second is to eliminate the variable
in the first potential in favor of the other variable to obtain the second
potential. The last step involves the derivative of the first potential, as
the Legendre transform is associated to an extremum value. But stopping the
procedure at the first step and use of the generalized potential that
depends on the two conjugate variables is not devoid of use. Familiar
examples of incomplete Legendre transforms are the Landau free energy (when
describing a magnet it has a dependence on both magnetization and external
field) and the free energy density functionals associated to many thermal
problems. Eq. (\ref{zipf2}), being the inverse of Eq. (\ref{benford2}),
states the same relationship but in terms of the `partition functions' $N(k)$
and $N_{\max }(\mathcal{N}^{-1})$. The absence of an upper bound for the
rank $k$ indicates a condition we refer to as the thermodynamic limit in our
statistical-mechanical interpretation of Eq. (\ref{benford2}). To complete
the Legendre transformation of $\widehat{S}_{\alpha }(\mathcal{N}^{-1})$
into $S_{\alpha }(k)$ and eliminate the variable $\mathcal{N}^{-1}$\ in
favor of $k$, it would be required to optimize $S_{\alpha }$, i.e. via the
use of an `equation of state'%
\begin{equation}
k=\frac{d}{d\mathcal{N}^{-1}}\log _{\alpha }N_{\max }(\mathcal{N}^{-1}).
\label{state1}
\end{equation}%
We address this issue in more detail in Section 5.

\section{Analogy with the tangent bifurcation}
\label{analogy}

Remarkably, there is a strict analogy between the generalized law of Zipf,
Eqs. (\ref{benford2}) and (\ref{zipf2}), and the nonlinear dynamics for the
RG fixed-point map at the tangent bifurcation, as originally realized in
Ref. \cite{hu1}. Consequently, these two apparently different problems share
the same statistical-mechanical interpretation indicated in the previous
section, and the equivalence offers an alternative to advance our analysis,
specifically, the characterization of finite size effects for the
generalized law in terms of the shift of the map out of tangency.

The analogy can be seen immediately after a brief recall of the RG treatment
of the tangent bifurcation that mediates the transition between chaotic and
periodic attractors \cite{schuster1}. The common procedure to study the
transition to chaos from a trajectory of period $n$ starts with the
composition $f^{(n)}(x)$ of a one-dimensional map $f(x)$ at such
bifurcation, followed by an expansion for the neighborhood of one of the $n$
points tangent to the line with unit slope \cite{schuster1}. With complete
generality one obtains

\begin{equation}
x^{\prime }=f^{(n)}(x)=x+ux^{z}+...,\;x\geq 0,\;z>1,  \label{tangent1}
\end{equation}%
where $x^{z}\equiv $sign$(x)\left\vert x\right\vert ^{z}$. The RG
fixed-point map is the solution $f^{\ast }(x)$ of%
\begin{equation}
f^{\ast }(f^{\ast }(x))=\lambda ^{-1}f^{\ast }(\lambda x)
\label{fixedpoint1}
\end{equation}%
together with a specific value for $\lambda $ that upon expansion around $x=0
$ reproduces Eq. (\ref{tangent1}). An exact analytical expression for $%
f^{\ast }(x)$ was obtained in Ref. \cite{hu1} with the use of the assumed
translation property of an auxiliary variable, $y=x^{1-z}$. This property is
written as

\begin{equation}
x^{\prime 1-z}=x^{1-z}+(1-z)u  \label{fixedpoint2}
\end{equation}%
or, equivalently, as

\begin{equation}
x^{\prime }=x\exp _{z}(ux^{z-1}).  \label{fixedpoint3}
\end{equation}%
It is straightforward to corroborate that $x^{\prime }=f^{\ast }(x)$ as
given by Eq. (\ref{fixedpoint3}) satisfies Eq. (\ref{fixedpoint1}) with $%
\lambda =2^{1/(z-1)}$. Repeated iteration of Eq. (\ref{fixedpoint2}) leads to%
\begin{equation}
x_{t}^{1-z}=x_{0}^{1-z}+(1-z)ut  \label{trajectory1}
\end{equation}%
or%
\begin{equation}
\log _{z}x_{t}=\log _{z}x_{0}+ut.  \label{trajectory2}
\end{equation}%
So that the iteration number or time $t$ dependence of all trajectories is
given by%
\begin{equation}
x_{t}=x_{0}\exp _{z}\left[ x_{0}^{z-1}ut\right],  \label{trajectory3}
\end{equation}%
where the $x_{0}$ are the initial positions. The $q$-deformed properties of
the tangent bifurcation are discussed at greater length in Ref. \cite%
{baldovin1}. The parallel between Eqs. (\ref{trajectory2}) and (\ref%
{trajectory3}) with Eqs. (\ref{benford2}) and (\ref{zipf2}), respectively,
is plain, and therefore we conclude that the dynamical system represented by
the fixed-point map $f^{\ast }(x)$ operates in accordance to the same
statistical-mechanical property described in the previous section for the
generalized laws.

To emphasize that there is a firm analogy, not a casual resemblance, between
the ranking of data and the sequences of iterates at the tangent bifurcation
we show that there is a common source behind Eqs. (\ref{benford2}) and (\ref%
{trajectory2}), i.e. the restriction of accessibility to phase space already
mentioned. This is readily seen by considering the replacement, valid for
large time $\tau $, of the difference $x_{\tau +1}-x_{\tau }$ by $dx_{\tau
}/d\tau $ in Eq. (\ref{tangent1}), written as $x_{\tau +1}-x_{\tau
}=u\left\vert x_{\tau }\right\vert ^{z}$. Integration of the left hand side
of the resulting differential form%
\begin{equation}
\frac{dx_{\tau }}{\left\vert x_{\tau }\right\vert ^{z}}=ud\tau
\label{differentialform1}
\end{equation}%
between $x_{0}$ and $x_{t}$ and the right hand side from $0$ to $t$ leads
immediately to Eqs. (\ref{trajectory1}) or (\ref{trajectory2}). The quantity 
$\left\vert x_{\tau }\right\vert ^{-z}$ in Eq. (\ref{differentialform1})
plays the same role as the power law distribution $P(N)\sim N^{-\alpha }$.

\begin{figure}[tbp]
\centering
\includegraphics[height=6.2cm]{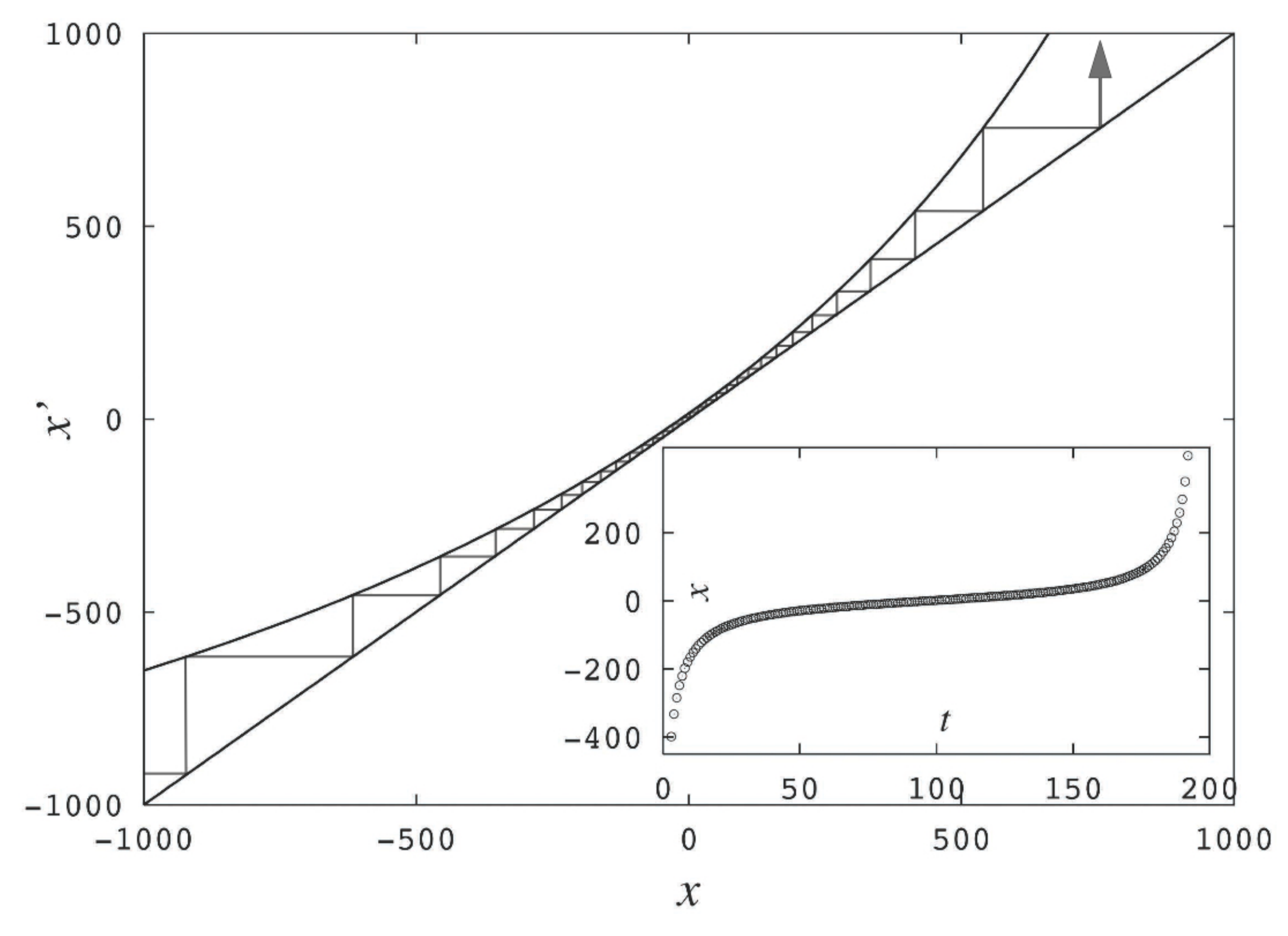}
\caption{The map in Eq.(\protect\ref{offtangency1}) with a trajectory. The
inset shows the time dependence of the trajectory.}
\label{fig:map}
\end{figure}

We notice that the absence of an upper bound for the rank $k$ in Eqs. (\ref%
{benford2}) and (\ref{zipf2}) is equivalent to the tangency condition in the
map. Accordingly, we look at the changes in $N(k)$ brought about by shifting
the corresponding map from tangency (see Fig.~\ref{fig:map}), i.e., we
consider the trajectories $x_{t}$ with initial positions $x_{0}$ of the map%
\begin{equation}
x^{\prime }=x\exp _{z}(ux^{z-1})+\varepsilon ,\;0<\varepsilon \ll 1
\label{offtangency1}
\end{equation}%
with the identifications $k=t$, $\mathcal{N}^{-1}=-u$, $N(k)=x_{t}+x^{\ast }$%
, $N_{\max }=x_{0}+x^{\ast }$ and $\alpha =z$, where the translation $%
x^{\ast }$ ensures that all $N(k)\geq 0$.

\begin{figure}[tbp]
\centering
\includegraphics[height=6.2cm]{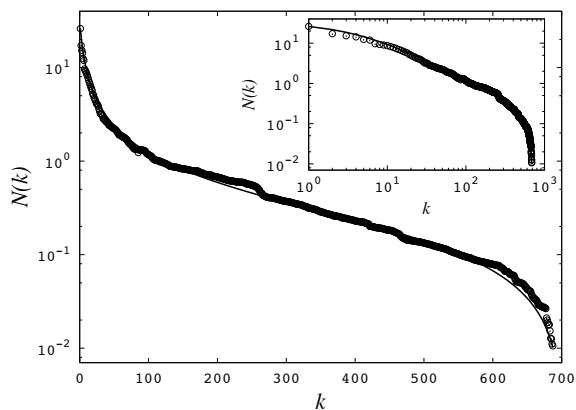}
\caption{Rank-order statistics for the eigenfactor of physics journals (%
\emph{empty circles}) from Ref. \protect\cite{eigenfactor1}. Eq. (\protect
\ref{offtangency1}) with the identifications provided in the text when $%
\protect\alpha =2.01$ and $\protect\epsilon =-0.00064$ (\emph{smooth curve})
is fitted to the data.}
\label{fig:eigen}
\end{figure}

\begin{figure}[tbp]
\centering
\includegraphics[height=6.2cm]{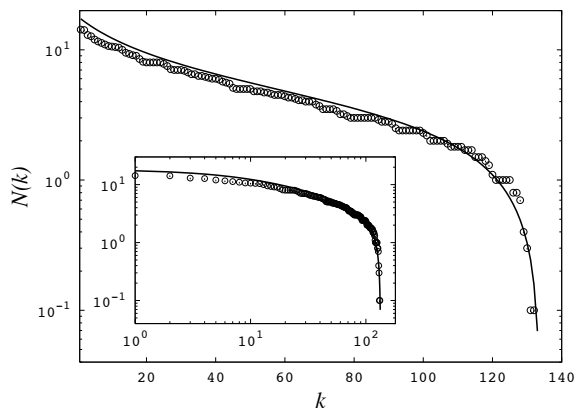}
\caption{Rank-order statistics for industrial production growth rates (\emph{%
empty circles}) from Ref. \protect\cite{industrial1}. Eq. (\protect\ref%
{offtangency1}) with the identifications provided in the text when $\protect%
\alpha =2.13$ and $\protect\epsilon =-0.058$ (\emph{smooth curve}) is fitted
to the data.}
\label{fig:industrial}
\end{figure}

\begin{figure}[tbp]
\centering
\includegraphics[height=6.2cm]{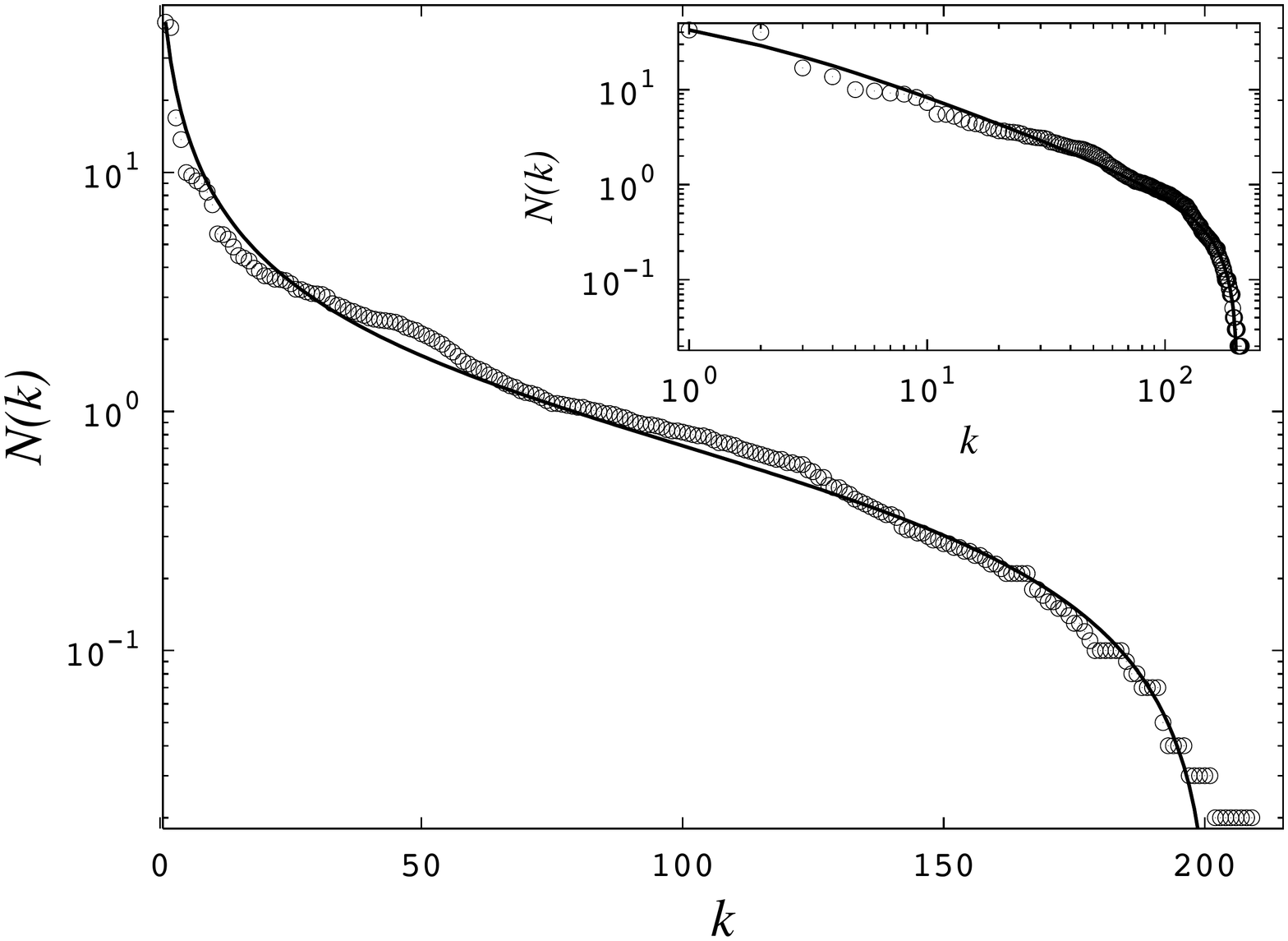}
\caption{Rank-order statistics for per capita carbon dioxide emissions (%
\emph{empty circles}) from Ref.\protect\cite{emissions1}. Eq. (\protect\ref%
{offtangency1}) with the identifications provided in the text when$\protect%
\alpha =2.06$ and $\protect\epsilon =-0.0055$ (\emph{smooth curve}) is
fitted to the data.}
\label{fig:carbon}
\end{figure}

In Figs.~\ref{fig:eigen} to~\ref{fig:carbon} we illustrate the capability of
this approach to reproduce quantitatively real data for ranking of
eigenfactors (a measure of the overall value) of physics journals \cite%
{eigenfactor1}, industrial production growth rates by country \cite%
{industrial1}, and per capita carbon dioxide emissions by country or region 
\cite{emissions1}, respectively.

In the intermittency route out of chaos it is relevant to determine the
duration of the so-called laminar episodes \cite{schuster1}, i.e., the
average time spent by the trajectories going through the \textquotedblleft
bottle neck" formed in the region where the map is closest to the line of
unit slope. Naturally, the duration of the laminar episodes diverges at the
tangent bifurcation when the Lyapunov exponent for separation of
trajectories vanishes. Interestingly, it is this property of the nonlinear
dynamics that translates into the finite-size ($k_{\max }<\infty $)
properties of the occurrence-rank function $N(k)$, that we have obtained
without finding out the details of the departure of the basic distribution $%
P(N)$ from the pure power-law $N^{-\alpha }$. One more important result that
follows from the analogy between nonlinear dynamics and the rank law is that
the most common value for the degree of nonlinearity at tangency is $z=2$,
obtained when the map is analytic at $x=0$ with nonzero second derivative,
and this implies $\alpha =2$, close to the values observed for most sets of
real data.

\section{Universality and uniqueness of data ranking}
\label{universality}

The steepest-descent approximation is central to statistical mechanics (and
in a more general context to large deviation theory \cite{touchette1}). This
property facilitates the evaluation in the thermodynamic limit of a
partition function for one particular ensemble in terms of the partition
function of another. Thermodynamically, this approximation relates to the
Legendre transformation between the corresponding free energies or Massieu
potentials where one variable is eliminated in favor of its conjugate \cite%
{callen1}. As recalled above, the procedure consists of two steps, summation
(or subtraction) of the product of the conjugate variables to (or from) the
first potential to define the second, followed by use of the derivative of
the first potential, or equation of state, to remove the undesired variable.
This, of course, corresponds to the optimization involved in the
steepest-descent method. For illustrative purposes we will assume here that
the steepest-descent shortcut that underlies the second step in the Legendre
transformation is also meaningful for $\alpha >1$.

In order to carry out the second step in the Legendre transform stated by
Eq. (\ref{legendre1}) we need an explicit form for the function $N_{\max }(%
\mathcal{N}^{-1})$. It is evident that the form of this function is not
unique and is determined by the particular set of data $\mathcal{N}$. For
illustrative purposes we consider a finite set of data $\mathcal{N}$
extracted from $P(N)=N^{-\alpha }$ although a pure power law is not the
correct distribution in this case. However, if the equivalent map Eq. (\ref%
{offtangency1}) is very close to tangency $\varepsilon \ll 1$ and the data
for the maximum rank, $N_{\min }=N(k_{\max })$, is chosen such that its
image in the map is to the left and near to its bottleneck mid-point, then $%
P(N)$ is closely approximated by the power law$N^{-\alpha }$. Under this
approximation normalization of $P(N)$ only yields $k_{\max }\sim $ $\mathcal{%
N}$. Suppose the available data, or the choice of the data collector, fixes
the specific value of $k_{\max }$ and the lowest and upper limits in Eq. (%
\ref{rank1}) to be $N_{\min }$ and $N_{\max }$, respectively. Therefore we
have%
\begin{equation}
k_{\max }=\mathcal{N}\int\limits_{N_{\min }}^{N_{\max }}N^{-\alpha }dN=\frac{%
\mathcal{N}}{1-\alpha }\left[ N_{\max }^{1-\alpha }-N_{\min }^{1-\alpha }%
\right] ,  \label{maxrank1}
\end{equation}%
$\alpha \neq 1$, or%
\begin{equation}
N_{\max }=N_{\min }\exp _{\alpha }[N_{\min }^{1-\alpha }\mathcal{N}%
^{-1}k_{\max }].  \label{Nmax1}
\end{equation}%
For example, a set of data about population of cities may be represented by $%
k_{\max }=50$ (fifty representative city sizes), $N_{\min }=1$ (one city
with the largest population), and $N_{\max }=100$ (one hundred cities with
the smallest population considered). Eq. (\ref{Nmax1}) is the required
expression for $N_{\max }(\mathcal{N}^{-1})$ to be used in the
`steepest-descent condition' or `equation of state' Eq. (\ref{state1}). The
result follows immediately, it is $k=$ $k_{\max }$.

As in ordinary thermodynamics, we observe that the universality of the laws
described by Eqs. (\ref{benford2}) and (\ref{zipf2}) is due to the general
form of the incomplete Legendre transformation, while the\ specific forms
adopted by the potentials $\widehat{S}_{\alpha }=\log _{\alpha }N_{\max }(%
\mathcal{N}^{-1})$ and $S_{\alpha }=\log _{\alpha }N(k)$ are particular to
the system or situation considered.

\section{Summary and discussion}
\label{summary}

We have suggested here a novel thermodynamic, or statistical-mechanical,
interpretation or understanding of the generalized laws of Benford and Zipf.
The expressions for these laws, Eqs. (\ref{benford1}) and (\ref{rank1}) (or
alternatively (\ref{zipf2})) were derived in Ref. \cite{pietronero1} under
the basic assumption that the data sets obeyed by these laws are
statistically well reproduced when extracted from a power law distribution $%
P(N)\sim N^{-\alpha }$. We remark here that the deviation from unity of the
exponent $\alpha $ implies a restricted access to the phase space for the
data configurations that when enumerated produce the numbers $N$. The
restriction involves an accessible subset of this space with a scale
invariant property, i.e., a fractal set, as implied by the power law $%
N^{-\alpha }$. This viewpoint becomes evident when $P(N)$ is seen to
represent the probability distribution of $N$ equally probable
configurations in the phase space for the data, and, consequently, suggests
the definition of the generalized entropies in Eq. (\ref{entropies1}). It is
important to clarify that the statistical-mechanical structure considered
here and obtained from the usual via a scalar deformation parameter
(represented by the power $\alpha $) does not conform to that known as
nonextensive statistics \cite{tsallis1} \cite{tsallis2}. Even though we
define entropies or Massieu potentials with the use of the $q$-logarithmic
function and make use of its inverse, the $q$-exponential, we do not require
or implicate the optimization of any of these quantities via the use of the
constraints employed in the nonextensive formalism or involve the use of the
so-called escort distributions \cite{tsallis2}.

The ranking of real data habitually shows deviations from the Zipf's
power-law regime both for small and large rank that can be clearly observed
in semi-log plots. As we have shown in Fig. 1 the generalized Zipf's law
given by Eq. (\ref{zipf2}) is capable of reproducing accurately the low rank
deviation but not that for large rank as the power-law regime in this
equation extends to $k\rightarrow \infty $. An upper bound for $k$ suggests
finite-size effects inherent in real data. We have captured the nature of
the upper bound for $k$ by first demonstrating a precise analogy between the
expression for the ranking laws, Eqs. (\ref{benford2}) and (\ref{zipf2}),
and those for the dynamics at the transition to chaos via intermittency (the
tangent bifurcation) in nonlinear maps of low dimensions. The finite-size
effects in the ranking of data are seen to correspond to the shift off
tangency in the map, so that the position of the upper bound for the rank $k$
is given by the duration of the laminar episodes of chaotic trajectories
near the transition to regular behavior. Interestingly, the
statistical-mechanical interpretation put forward for the generalized law of
Zipf extends over to the critical dynamics of the transition to chaos via
intermittency. While, on the practical side, data for the ranking of data
for all $k$ is reproduced quantitatively by our formalism, as illustrated,
respectively, in Figs.~\ref{fig:eigen} to~\ref{fig:carbon} for three
specific examples: eigenfactor of physics journals \cite{eigenfactor1},
industrial production rates \cite{industrial1}, and carbon emissions \cite%
{emissions1}. In agreement with empirical determinations the analogy implies
that the most general value for the index $\alpha $ is $\alpha =2$.

As it is generally well-known, a statistical-mechanical structure (shared by
large deviation theory \cite{touchette1}) is built around the
steepest-descent approximation and is expressed as the Legendre transform
property that links different thermodynamic potentials. It involves an
optimization condition or equation of state that relates conjugate
variables. Only for illustrative purposes we have assumed that this
structure extends to the deformed version (with one scalar parameter) we
have considered here. In order to replicate the circumstances normally
encountered in thermodynamics we have presented as an example the particular
form taken by the function $N_{\max }(\mathcal{N}^{-1})$ when the data in
hand is bounded by the numbers $N_{\min }$ and $N_{\max }$ and fixes the
largest rank $k_{\max }$. Then, the equation of state was determined and the
variable $\mathcal{N}^{-1}$ eliminated in favor of $k$, to obtain the
`equilibrium' value for $N(k)$. This exercise suggests that the universality
of the laws is due to the general form of the incomplete Legendre
transformation, while the expressions for the initial and transformed
potentials are specific to the design of the data sample under
consideration. The thermodynamic interpretation we have put forward may
explain the ever presence of these phenomenological laws in a wide range of
observations including very dissimilar situations. Finally, we comment that
our arguments also apply to the topic of scale-free networks \cite{barabasi1}%
. Since the degree distribution $p(k)$, the distribution for the number $k$
of links that connect one node to other nodes, describes essentially the
ranking of nodes according to the number of links they possess, we can treat
the data sets from where this distribution is phenomenologically obtained
similarly to the data sets leading to Zipf's law. Interestingly, for random
link networks $p(k)$ decays exponentially ($\alpha =1$), but for scale-free
networks it is approximately power law ($\alpha >1$).

\section*{Acknowledgements}

We recognize support from DGAPA-UNAM and CONACyT (Mexican agencies) and MEC
(Spain). A.R. is grateful to the Grupo Interdisciplinar de Sistemas
Complejos (GISC) for hospitality in Madrid.

\end{document}